\documentclass{amsart}

\usepackage{hyperref}
\usepackage{pgfplotstable}
\usepackage{booktabs}
\usepackage{array}
\usepackage{colortbl}
\usepackage{mathtools}
\usepackage{graphicx}
\usepackage{figsize}
\usepackage{enumerate}
\usepackage{enumitem} 
\usepackage{amsmath,amssymb,latexsym, geometry}

\usepackage{color}
\usepackage{subfigure}
\usepackage{multirow}
\usepackage{mathptmx}
\usepackage{url,bm,overpic}
\usepackage{enumerate}
\usepackage{bbm}
\usepackage[ruled,vlined,linesnumbered]{algorithm2e}
\usepackage{comment}
\newtheorem{theorem}{Theorem}[section]

\theoremstyle{remark}
\newtheorem{remark}{Remark}[section]

\numberwithin{equation}{section}

\pgfplotsset{compat=1.18}

\begin{document}

\begingroup
\def\uppercasenonmath#1{}
\let\MakeUppercase\relax

\title{{\Large Scattering and inverse scattering for multipoint potentials at high energies}}

\author{}
\maketitle

\markleft{}
\markright{}
\pagestyle{plain}  
\vspace{-2ex}
\begin{center}
    { \textbf{P.C. Kuo}}\quad \textbf{and} \quad \textbf{R.G. Novikov}
\end{center}

\begin{abstract}
We consider the Schrödinger equation with a multipoint potential of Bethe-Peierls-Thomas-Fermi type. For this singular potential, we develop scattering and inverse scattering at high energies. In particular, in this framework, our results include analogs of the "regular" Born-Faddeev formula for the scattering amplitude and analogs of related "regular" inverse scattering reconstructions at high energies. Related results for scattering solutions at high energies are also presented.
\end{abstract}



\section{Introduction}
\label{intro}
We consider the Schrödinger equation

\begin{equation}
    -\Delta \psi+v(x) \psi=E \psi, \quad x \in \mathbb{R}^{d}, \quad E>0,
    \label{Schrodinger}
\end{equation}
where $v(x)$ decays sufficiently fast as $|x| \to \infty$. 

For this equation, we consider the scattering solutions $\psi^{+}$ such that
\begin{equation} 
\label{wave}
\psi^{+}(x,k)=e^{i k \cdot x} +\psi^{sc}(x,k),\quad x, k \in \mathbb{R}^{d},\quad |k|=\sqrt{E},
\end{equation}
where $\psi^{sc}(x,k)$ satisfies the Sommerfeld radiation condition:
\begin{equation}
\label{Sommerfeld}
|x|^{\frac{d-1}{2}} \left( \frac{\partial}{\partial |x|} - i\kappa\right) \psi^{sc}(x,k) \xrightarrow[|x| \to \infty]{} 0 \quad \text{uniformly in all directions } \frac{x}{|x|},
\end{equation}
where $\kappa = \sqrt{E}>0$.
Then,
\begin{equation} 
\label{f+}
\psi^{sc}(x,k)= \frac{e^{i\kappa|x|}}{|x|^{(d-1) / 2}} f^{+}(k, \frac{\kappa}{|x|}x)+O\left(\frac{1}{|x|^{(d+1) / 2}}\right) \quad \text{as } \quad |x| \rightarrow \infty.
\end{equation}
The function $f^{+}$ arising in (\ref{f+}) is the scattering amplitude for equation (\ref{Schrodinger}). This function is defined on $\mathcal{M}_E$, where
\begin{equation}
\label{sphere}
\mathcal{M}_E=\{\,k,l\in\mathbb{R}^d:\ k^2=l^2=E\,\}
=\mathbb{S}^{d-1}_{\kappa}\times \mathbb{S}^{d-1}_{\kappa},
\end{equation}
\begin{equation*}
     \mathbb{S}_{r}^{d-1}=\{m \in \mathbb{R}^{d}:\ |m|=r\}.
\end{equation*}

In connection with definitions and properties of the aforementioned functions $\psi^{+}$ and $f^{+}$, see, for example, \cite{faddeev1956uniqueness}, \cite{novikov2019multidimensional}, \cite{yafaev2003high}, for the case of regular $v$, and \cite{albeverio2012solvable}, \cite{demkov2013zero}, \cite{kuo2024inverse}, for the singular case of multipoint $v$ of Bethe-Peierls-Thomas-Fermi type, and references therein. 

In addition, the problem of finding $f^{+}$ and $\psi^{+}$ from $v$ is known as a problem of direct scattering, whereas the problem of reconstructing $v$ from some information about $f^{+}$ or $\psi^{+}$ is known as a problem of inverse scattering. These problems also include studies of different properties of $f^{+}$ and $\psi^{+}$. 

Many important results on scattering and inverse scattering for equation (\ref{Schrodinger}) with regular $v$ are given in the literature; see, for example, \cite{novikov2019multidimensional}, \cite{yafaev2003high} and references therein. In particular, the following Born-Faddeev formula holds: 
\begin{equation}
\label{Born-Faddeev}
f(k,\ell)=\hat{v}(k-\ell)+O\!\left(E^{-1/2}\right),\quad E \to +\infty,\quad (k,\ell)\in \mathcal{M}_E,
\end{equation}
where
\begin{equation}
f^{+}(k,\ell)=c(d,\kappa)f(k,\ell);\quad c(d,\kappa) =-\pi i(\sqrt{2 \pi} e^{-i \pi / 4} )^{(d-1)}\kappa^{(d-3) / 2},  
\end{equation}
and 
\begin{equation}
\label{Fourier}
\hat{v}(p)= \mathcal{F}v(p)=(2\pi)^{-d}\int_{\mathbb{R}^d} e^{ipx}\,v(x)\,dx,\quad p\in\mathbb{R}^d.
\end{equation}
Formula (\ref{Born-Faddeev}) goes back to \cite{born1926quantenmechanik}, \cite{faddeev1956uniqueness}. This formula holds, for example, for compactly supported $v \in L^{\infty}(\mathbb{R}^{d})$. In addition, for infinitely smooth $v$ with sufficient decay at infinity, the scattering amplitude $f$ admits a high-energy asymptotic expansion containing all terms of power order with respect to $\kappa^{-n}$; see \cite{buslaev1967trace}, \cite{yafaev2003high} and references therein.

In dimension $d\geq 2$, formula (\ref{Born-Faddeev}) gives the simplest method for finding $v$ from $f$ at high energies for the case of regular $v$; see \cite{faddeev1956uniqueness}, \cite{novikov2019multidimensional} and references therein.

Besides, for sufficiently regular $v$ with sufficient decay at infinity, the following high-energy Born-type formula for $\psi^{sc}$ in (\ref{wave}) is also well-known:
\begin{equation}
\label{Yafayev}
    \psi^{sc}(x,k) =  e^{ik\cdot x} (2i\kappa)^{-1}\, \mathcal{D}v(x,-\theta) + \mathcal{O}(\kappa^{-2}),\quad  E \to +\infty, \quad \text{at fixed } x \text{ and }  \theta = k/\kappa,
\end{equation}
where 
\begin{equation}
\label{beam transform}
   \mathcal{D}v(x,\theta) = \int_{0}^{\infty} v(x + t\theta)\, dt, \quad x\in \mathbb{R}^{d}, \quad \theta \in \mathbb{S}^{d-1},
\end{equation}
i.e., $\mathcal{D}v$ is the divergent beam transform of $v$. In addition, for infinitely smooth $v$ with sufficient decay at infinity, the scattering solutions $\psi^{+}$ admit a high-energy asymptotic expansion containing all terms of power order with respect to $\kappa^{-n}$. In connection with these asymptotic results for $\psi^{+}$, see \cite{buslaev1967trace}, \cite{yafaev2003high} and references therein.

Besides, in dimension $d \geq 2$, formula (\ref{Yafayev}) gives a method for finding $v$ from high-energy boundary values of $\psi^{sc}$ on $\partial \Omega$, for the case when 
$\mathrm{supp}\, v \subset \Omega$, $\Omega$ is a bounded domain in $\mathbb{R}^{d}$. This method involves finding $v$ from $\mathcal{P}v$, where $\mathcal{P}v$ is the X-ray transform of $v$, and uses relations between $\mathcal{P}v$ and $\mathcal{D}v$; see, e.g., \cite{natterer2001mathematics} for introduction into mathematics of these transforms. Note also that results on reconstructing $v$
from boundary values of $\psi^{sc}$ go back to \cite{ie1957behavior} (which is already cited in \cite{faddeev1956uniqueness}).

The main purpose of the present work consists in presenting an analog of formula (\ref{Born-Faddeev}) and an analog of the aforementioned inverse scattering at high energies for the case of multipoint potential $v$ of Bethe–Peierls–Thomas-Fermi type:
 \begin{equation}
     \quad v(x)=\sum\limits_{j=1}^{n} \delta_{\alpha_{j}}\left(x-y_{j}\right), \quad \alpha_{j} \in \mathbb{C}, \quad y_{j} \in \mathbb{R}^{d}, \quad y_{j} \neq y_{j'} \quad \forall j \neq j', \quad n \in \mathbb{N} \cup\{0\},
     \label{multipotential}
 \end{equation}
where $d=1,2,3$, and $v \equiv 0$ for $n=0$.

Here, $\delta_{\alpha}(x)=\epsilon \delta(x)$, where $\epsilon=-1/\alpha$ and $\delta$ is the standard Dirac delta function for $d=1$, whereas $\delta_{\alpha}$ denotes a "renormalized" delta function for $d=2$ and $d=3$; see \cite{kuo2024inverse} and references therein for precise definition of $\delta_{\alpha}$ for the latter cases. Historically, the point scatterers corresponding to $\delta_{\alpha}$ in dimension $d=3$ were first introduced to describe the interaction between neutrons and protons by Bethe, Peierls \cite{bethe1935quantum}, Thomas \cite{thomas1935interaction}, and Fermi \cite{fermi1936sul}. Scatterers with similar mathematical properties also arise, in particular, in acoustics; see \cite{badalyan2009scattering}, \cite{dmitriev2021features} and references therein. For detailed information about multipoint scatterers $v$ in (\ref{multipotential}), see, for example, \cite{albeverio2012solvable}, \cite{demkov2013zero}, \cite{grinevich2021transmission}. In particular, it is known that such multipoint scatterers $v$, for $n \geqslant1$, are only defined in dimensions $d=1,2,3$.

In the present work, we proceed from formulas on direct scattering for finding the scattering functions $\psi^{+}$ and $f$ from $y_{j}$'s and $\alpha_{j}$'s arising in (\ref{multipotential}); see formulas (\ref{eigenfunctions})-(\ref{coefficientA}) in Section 2.

The main results of the present work are given by Theorems \ref{main asymptotics} -- \ref{high-energy-asymptotics}, Remark \ref{rk3.3}, and Theorems \ref{psi-asymptotics}, \ref{psi-beam transform} in Section \ref{Main}.

In particular, Theorem \ref{main asymptotics} gives analogs of formula (\ref{Born-Faddeev}) for the case of multipoint potentials in (\ref{multipotential}) for $d=2,3$. As a corollary, in Theorem \ref{thm:det}, we present a reconstruction procedure for high-energy inverse scattering for $v$ in (\ref{multipotential}) for $d=2,3$. In turn, Theorem \ref{main asymptotics} follows from Theorem \ref{high-energy-asymptotics}, containing full high-energy asymptotic expansions of the scattering amplitude $f$ for $v$ in (\ref{multipotential}) for $d=1,2,3$.

In addition, Remark \ref{rk3.3} and Theorems \ref{psi-asymptotics}, \ref{psi-beam transform} give high-energy asymptotics for scattering solutions $\psi^{+}$ for the case of multipoint potentials in (\ref{multipotential}).

By the aforementioned results, we contribute to scattering and inverse scattering for model (\ref{Schrodinger}), (\ref{multipotential}) at high energies. In connection with previous results on direct scattering and spectral theory for this model, see, for example, \cite{albeverio2012solvable}, \cite{Berezin:1960df}, \cite{demkov2013zero}, \cite{grinevich2022spectral}, \cite{grinevich2024transparent}, \cite{malamud2024kernels} and references therein. In connection with previous results on inverse scattering for this model, see \cite{agaltsov2019examples}, \cite{badalyan2009scattering}, \cite{dmitriev2021features}, \cite{gesztesy1999inverse}, \cite{grinevich2012faddeev}, \cite{grinevich2021transmission}, \cite{kuo2024inverse}, \cite{mantile2023inverse}, \cite{novikov2018inverse} and references therein.

The main results of this work are presented in detail in Section 3. Proofs are given in Section 4.

\section{Preliminaries}

 Let $G^{+}=G^{+}(x,\kappa)$ be the  Green function with the Sommerfeld radiation condition for the operator $\Delta+E$, where $E=\kappa^2, \kappa>0$, that is 
 
\begin{align}
G^{+}(x, \kappa)&=-(2 \pi)^{-d} \int_{\mathbb{R}^{d}} \frac{e^{i \xi \cdot x} d \xi}{|\xi|^{2}-\kappa^2-i \cdot 0}    \label{Green} \\
&= 
\begin{cases}
  \frac{e^{i\kappa|x|}}{2 i\kappa}  &  ,\text{if } d=1  \\
  -\frac{i}{4} H_{0}^{1}(|x|\kappa) & ,\text{if }d=2 \\
  -\frac{e^{i\kappa|x|}}{4 \pi|x|} & ,\text{if }d=3 \\ 
\end{cases} \nonumber
\end{align}
where $H_{0}^{1}$ denotes the zeroth-order Hankel function of the first kind. Note that $G^{+}(x,\kappa)$ only depends on $|x|$ when $\kappa$ is fixed.

If $v$ is a multipoint potential of Bethe–Peierls–Thomas-Fermi type (\ref{multipotential}), then the following explicit formulas for the scattering functions $\psi^{+}$ and $f$ hold (see \cite{grinevich2024transparent}, \cite{kuo2024inverse} and references therein):

\begin{align}
\psi^{+}(x, k) &=e^{i k \cdot x}+\sum_{j=1}^{n} q_{j}(k) G^{+}\left(x-y_{j},\kappa\right), \quad x \in \mathbb{R}^{d},\quad k \in \mathbb{S}^{d-1}_{\kappa}, \label{eigenfunctions} \\
f(k, l) &=\frac{1}{(2 \pi)^{d}} \sum_{j=1}^{n} q_{j}(k) e^{-i l \cdot y_{j}},\quad k,l \in \mathbb{S}^{d-1}_{\kappa},  \label{scattering amplitude}
\end{align}

where $ \mathbb{S}^{d-1}_{\kappa}$ is defined in (\ref{sphere}) and $q(k)=\left(q_{1}(k), \cdots, q_{n}(k)\right)^{t}$ satisfies
\begin{align}
    A(\kappa) q(k) &=b(k), \label{A=qb} \\
    A_{j, j}  (\kappa) &= 
    \begin{cases}\alpha_{j}+(2 i \kappa)^{-1} &, \text {if } d=1 \\ \alpha_{j}-(4 \pi)^{-1}(\pi i-2 \ln (\kappa)) &, \text {if } d=2 \\
    \alpha_{j}-i(4 \pi)^{-1}\kappa &, \text {if } d=3 \\
    \end{cases} \label{coefficientA} \\
    A_{j, j^{\prime}}(\kappa) &= G^{+}\left(y_{j}-y_{j^{\prime}}, \kappa \right), j \neq j^{\prime}, \nonumber \\
    b(k) &=-\left(e^{i k \cdot y_{1}}, e^{i k \cdot y_{2}}, \cdots e^{i k \cdot y_{n}}\right)^{t}. \nonumber
\end{align}

For more details on direct scattering for multipoint potentials in (\ref{multipotential}) and their possible modifications, see, e.g., \cite{albeverio2012solvable}, \cite{badalyan2009scattering}, \cite{chashchin2018example}, \cite{demkov2013zero}, \cite{dmitriev2021features}, \cite{grinevich2024transparent}, \cite{kuo2024inverse}, \cite{kurasov2009triplet}, \cite{loran2022renormalization}, \cite{malamud2024kernels}.

\section{Main results}\label{Main}

For multipoint potentials $v$ of Bethe-Peierls-Thomas-Fermi type, analogs of the Born-Faddeev formula (\ref{Born-Faddeev}) are given in the next theorem.

\begin{theorem}[]
\label{main asymptotics}
Let $f$ be the scattering amplitude for a multipoint potential $v$ of the type (\ref{multipotential}). Then the following asymptotics formulas hold as $\kappa := \sqrt{E} \to +\infty$:

\medskip
If $d=1$, then
  \begin{equation}
  \label{main1}
    f(k,\ell)
    = \sum_{j=1}^{n}
      \frac{-1}{2\pi \alpha_{j}}\,
      e^{i (k-\ell)\cdot y_{j}}
    + \mathcal{O}(\kappa^{-1});
  \end{equation}

If $d=2$, then
  \begin{align}
  \label{main2}
    f(k,\ell)
    &= (\ln \kappa)^{-1}
      \Biggl[
        \sum_{j=1}^{n}
          \frac{-1}{2\pi}
          e^{i (k-\ell)\cdot y_{j}}
      \Biggr] + (\ln \kappa)^{-2}
      \Biggl[
        \sum_{j=1}^{n}
          \left(\alpha_{j}-\tfrac{i}{4}\right)
          e^{i (k-\ell)\cdot y_{j}}
      \Biggr]
      + \mathcal{O}\bigl((\ln \kappa)^{-3}\bigr);
  \end{align}

If $d=3$, then
  \begin{align}
  \label{main3}
    f(k,\ell)
    &= \frac{1}{\kappa}
       \sum_{j=1}^{n}
       \frac{-i}{2\pi^{2}}e^{i (k-\ell)\cdot y_{j}} + \frac{1}{\kappa^{2}}\Biggl[
       -\sum_{j=1}^{n}
         \frac{2 \alpha_{j}}{\pi}\,
         e^{i (k-\ell)\cdot y_{j}}
       +
       \sum_{j\neq j'}
         \frac{e^{i \kappa |y_{j}-y_{j'}|}}
              {2\pi^{2}|y_{j}-y_{j'}|}
         e^{i (k\cdot y_{j} - \ell\cdot y_{j'})}\Biggr]
       + \mathcal{O}(\kappa^{-3}).
  \end{align}
  
Here, $(k,\ell) \in \mathcal{M}_E$, where $\mathcal{M}_E$ is defined in (\ref{sphere}).
\end{theorem}

To our knowledge, formulas (\ref{main2}) and (\ref{main3}) are new, whereas formula (\ref{main1}) is a particular case of formula
(\ref{Born-Faddeev}). In addition, each of the formulas (\ref{main1}), (\ref{main2}), (\ref{main3}) has its own specific structure.

Theorem \ref{main asymptotics} follows from Theorem \ref{high-energy-asymptotics} given below.

For multipoint potentials $v$ of Bethe-Peierls-Thomas-Fermi type, inverse scattering results based on Theorem \ref{main asymptotics} are given in Theorem \ref{thm:det} below.

Let us use the following notation for the leading coefficients in formulas (\ref{main1})-(\ref{main3}): 
\begin{equation}
\label{f1}
    f_{1,1}=\sum\limits_{j=1}^{n}
      \frac{-1}{2\pi \alpha_{j}}\,
      e^{i (k-\ell)\cdot y_{j}}, \quad d=1,
\end{equation}

\begin{equation}
\label{f2}
    f_{2,1}= 
        \sum_{j=1}^{n}
          \frac{-1}{2\pi}
          e^{i (k-\ell)\cdot y_{j}}, \quad
    f_{2,2}=
        \sum_{j=1}^{n}
          \left(\alpha_{j}-\tfrac{i}{4}\right)
          e^{i (k-\ell)\cdot y_{j}}, \quad d=2,
\end{equation}

\begin{align}
\nonumber
    \label{f3}
    f_{3,1}&=
       \sum_{j=1}^{n}
       \frac{-i}{2\pi^{2}}e^{i (k-\ell)\cdot y_{j}}, \quad 
    f_{3,2}(k-\ell)= f_{3,2}^{1}+ f_{3,2}^{2}, \\
    f_{3,2}^{1}&=
       -\sum_{j=1}^{n}
         \frac{2 \alpha_{j}}{\pi}\,
         e^{i (k-\ell)\cdot y_{j}}, \quad
    f_{3,2}^{2}=
       \sum_{j\neq j'}
         \frac{e^{i \kappa |y_{j}-y_{j'}|}}
              {2\pi^{2}|y_{j}-y_{j'}|}
         e^{i (k\cdot y_{j} - \ell\cdot y_{j'})},
         \quad d=3,
\end{align}
where $(k,\ell) \in \mathcal{M}_{E}$.

One can see that 
\begin{align}
    f_{1,1}=f_{1,1}(k,\ell)=\mathcal{F}v_{1,1}(k-\ell), \label{v1} \\
    f_{2,1}=f_{2,1}(k,\ell)=\mathcal{F}v_{2,1}(k-\ell), \label{v21}\\
    f_{2,2}=f_{2,2}(k,\ell)=\mathcal{F}v_{2,2}(k-\ell), \label{v22}\\
    f_{3,1}=f_{3,1}(k,\ell)=\mathcal{F}v_{3,1}(k-\ell), \label{v31} \\
    f_{3,2}^{1}=f_{3,2}^{1}(k,\ell)=\mathcal{F}v_{3,2}^{1}(k-\ell),\label{v321}
\end{align}
where $\mathcal{F}$ denotes the Fourier transform defined in (\ref{Fourier}), and
\begin{align}
    v_{1,1}(x)=v(x)= \sum\limits_{j=1}^{n}
      \frac{-1}{\alpha_{j}}\delta(x-y_{j}),\quad x \in \mathbb{R}, \label{delta11}
      \\
    v_{2,1}(x)= -2\pi\sum\limits_{j=1}^{n}
      \delta(x-y_{j}),\quad x \in \mathbb{R}^{2}, \label{delta21} \\
    v_{2,2}(x)= 4\pi^2 \sum\limits_{j=1}^{n}
      (\alpha_{j}-\frac{i}{4})\delta(x-y_{j}), \quad x \in \mathbb{R}^{2}, \label{delta22}\\
    v_{3,1}(x)=-4\pi i\sum\limits_{j=1}^{n}
      \delta(x-y_{j}),\quad x \in \mathbb{R}^{3}, \label{delta31}\\
    v_{3,2}^{1}(x)= -16 \pi^2 \sum\limits_{j=1}^{n} \alpha_{j}
      \delta(x-y_{j}),\quad x \in \mathbb{R}^{3}, \label{delta321}
\end{align}
where $\delta$ is the standard Dirac delta function.

In the next result, we also use the following parameterization for $(k,\ell)$ in formulas (\ref{main1})-(\ref{v321}):
\begin{equation}
k = k_E(p) = \frac{p}{2} + m_{E}(p), \quad
l = l_E(p) = -\frac{p}{2} + m_{E}(p), \quad p \in \mathbb{R}^{d}, \quad |p| < 2\sqrt{E}, 
\label{para}
\end{equation}
where
$$
m_E(p) = \left(E - \frac{p^2}{4}\right)^{1/2}\gamma(p), 
\quad |\gamma(p)| = 1, \quad \gamma(p)p = 0,
$$
where the square root is positive, $\gamma$ is a fixed vector-function, and $d \geq 2$.
\begin{theorem}
\label{thm:det}
Let $f$ be the scattering amplitude for a multipoint potential $v$ of the type (\ref{multipotential}). Then $f$ at high energies (i.e., for all energies $E>E_{0}\,$ for arbitrary fixed $E_{0}>0$) uniquely determines $v$ as follows:
\begin{itemize}
    \item Let $d=2$. Then $f(k_{E}(p), l_{E}(p))$ at high energies uniquely determines $\mathcal{F}v_{2,1}(p)$ and  $\mathcal{F}v_{2,2}(p)$ for each fixed $p \in \mathbb{R}^{2}$ via formulas (\ref{main2}),(\ref{f2}),(\ref{v21}),(\ref{v22}), (\ref{para}). In turn, $\mathcal{F}v_{2,j}$ uniquely determines $v_{2,j}$ for each $j=1,2$. Therefore, $\mathcal{F}v_{2,1}$ uniquely determines all $y_{j}$'s, whereas even $\mathcal{F}v_{2,2}$ alone uniquely determines all $y_{j}$'s and $\alpha_{j}$'s. This completes the determination of $v$.
    
    \item Let $d=3$. Then $f(k_{E}(p), l_{E}(p))$ at high energies uniquely determines $\mathcal{F}v_{3,1}(p)$ for each fixed $p \in \mathbb{R}^{3}$ via formulas (\ref{main3}), (\ref{f3}), (\ref{v31}), (\ref{para}). In turn, $\mathcal{F}v_{3,1}$ uniquely determines $v_{3,1}$, or, equivalently, all $y_{j}$'s.
    In addition, $y_{j}$'s uniquely determine $f_{3,2}^{2}$ in (\ref{f3}). Then $f(k_{E}(p), l_{E}(p))$ at high energies, $\mathcal{F}v_{3,1}(p)$, and $y_{j}$'s uniquely determine $\mathcal{F}v_{3,2}^{1}(p)$ for each fixed $p \in \mathbb{R}^{3}$  via formulas (\ref{main3}), (\ref{f3}), (\ref{v31}), (\ref{v321}), (\ref{para}). In turn, $\mathcal{F}v_{3,2}^{1}$ uniquely determines $v_{3,2}^{1}$, or, equivalently, all $y_{j}$'s and $\alpha_{j}$'s. This completes the determination of $v$.
    \item Let $d=1$.  Then $f(k, -k)$ at high energies uniquely determines $v$ via the formula \begin{equation}
\label{Ls}
    v(x)-\int_{|p|>2 \sqrt{E_{0}}} e^{-ipx}  f (\frac{p}{2},-\frac{p}{2})\, dp \in L^{s}(\mathbb{R}),\quad 2 \leq s < \infty,
\end{equation}
for sufficiently large $E_{0}$.
\end{itemize}

\end{theorem}

For $d=2,3$, Theorem \ref{thm:det} follows from (\ref{main2}), (\ref{main3}), (\ref{f2}), (\ref{f3}), (\ref{v21})-(\ref{v321}), (\ref{delta21})-(\ref{para}) and results on reconstruction of $v$ from $\mathcal{F}v$, where
\begin{equation}
    v(x)=\sum_{j=1}^{n} c_{j} \delta (x-y_{j}), \quad c_{j} \in \mathbb{C}\setminus\{0\}, \quad y_{j} \in \mathbb{R}^{d}, \quad y_{j} \neq y_{j'} \quad \forall j \neq j', \quad n \in \mathbb{N}\cup\{0\},
    \label{Helmholtz source}
\end{equation}
where $v\equiv 0$ for $n = 0$ and $\delta$ is the standard Dirac function.

For $d=1$, Theorem \ref{thm:det} follows from formulas (\ref{main1}), (\ref{f1}), (\ref{v1}), (\ref{delta11}). More precisely, in view of these formulas, we have that 
\begin{equation}
    \hat{v}(p) - \chi(p) f(\frac{p}{2},-\frac{p}{2}) \in L^{r} (\mathbb{R}), \quad  1 < r \leq 2,
\label{fourier,d=1}
\end{equation}
where $\chi (p) = 0 $ for $|p| \leq 2\sqrt{E_{0}}$ and  $\chi (p) = 1 $ for $|p| > 2\sqrt{E_{0}}$.
Formula (\ref{Ls}) follows from formula (\ref{fourier,d=1}) and the fact that the Fourier transform is bounded from $L^{r}(\mathbb{R})$ to $L^{s}(\mathbb{R})$, where $1<r\leq 2, \, 1/r + 1/s=1$. Note that formula (\ref{Ls}) is similar to Theorem 1, for $n=0$, in \cite{novikov1996inverse}, but we do not assume $v$ to be real in (\ref{Ls}). Note also that the condition that $E_{0}$ is sufficiently large in (\ref{Ls}) is necessary only for unique solvability of system (\ref{A=qb}) for $\kappa \geq \sqrt{E_{0}}$.

\begin{remark}
The result that a multipoint potential $v$ in (\ref{multipotential}), for $d=2,3$, is uniquely determined by its scattering amplitude $f$ at fixed energy is given in \cite{kuo2024inverse}. However, this determination in \cite{kuo2024inverse} involves analytic continuations and, therefore, is not appropriate for numerical implementations. The point is that reconstructions based on Theorem \ref{main asymptotics} and Theorem \ref{thm:det} already admit numerical implementations. Related numerical studies will be given elsewhere.
\label{rk3.1}
\end{remark}

Next, in addition to Theorem \ref{main asymptotics}, we also give full high-energy asymptotic expansions for the scattering amplitudes $f$ for multipoint potentials $v$ in (\ref{multipotential}).

\begin{theorem}[]\label{high-energy-asymptotics}
Let $f$ be the scattering amplitude for a multipoint potential $v$ of the type (\ref{multipotential}). Then the following full asymptotics formulas hold as $\kappa := \sqrt{E} \to +\infty$:

\noindent\textbf{(i)}  If $d=1$, then
\begin{equation}\label{eq:asymp-d1}
  f(k,\ell) 
  = \frac{-1}{2\pi}\sum_{m=0}^{M}   
      C_{m}(k,\ell)\,\kappa^{-m}
    + \mathcal{O}(\kappa^{-M-1}),
\end{equation}
where
\begin{equation}\label{eq:Pm-d1}
  C_{m}(k,\ell)= \sum_{j=1}^{n}\sum_{j'=1}^{n}
       [W^{m}]_{j,j'}\,\alpha_{j'}^{-1}\,
       e^{i(k\cdot y_{j'}- \ell\cdot y_{j})}, \quad  W_{j,j'}=\left(\frac{i}{2}\right)\alpha_{j}^{-1}e^{i \kappa |y_{j}-y_{j'}|},
\end{equation}
where $W^{m}$ is the $m$-th power of the matrix $W$.

In particular,
\begin{equation}\label{eq:P0-d1}
  \frac{-1}{2\pi}C_{0}(k,\ell)=f_{1,1}(k,\ell),
\end{equation}
where $f_{1,1}$ is defined in (\ref{f1}).

\medskip
\noindent\textbf{(ii) } If $d=2$, then
\begin{equation}\label{eq:asymp-d2}
  f(k,\ell)
  =  -(2 \pi \ln \kappa)^{-1}\sum_{m=0}^{M}
      (-2\pi)^{m}\,(\ln \kappa)^{-m}\,
      C_{m}(k,\ell)
    + \mathcal{O}\bigl((\ln \kappa)^{-M-2}\bigr),
\end{equation}
where
\begin{equation}\label{eq:Pm-d2}
  C_{m}(k,\ell)
  = \sum_{j=1}^{n}
    \left(\alpha_{j} - \tfrac{i}{4}\right)^{m}
    e^{i (k-\ell)\cdot y_{j}}
\end{equation}

In particular,
\begin{equation}\label{eq:P0-d2}
  \frac{-1}{2\pi}C_{0}(k,\ell)=f_{2,1}(k,\ell),\quad C_{1}(k,\ell)=f_{2,2}(k,\ell),
\end{equation}
where $f_{2,1}$, $f_{2,2}$ are defined in (\ref{f2}).

\medskip
\noindent\textbf{(iii)} If $d=3$, then
\begin{equation}\label{eq:asymp-d3}
  f(k,\ell)
  = -\frac{i}{2\pi^2}\,\kappa^{-1}\sum_{m=0}^{M}
      (-i 4\pi)^{m}
      C_{m}(k,\ell)\,
      \kappa^{-m-1}
    + \mathcal{O}(\kappa^{-M-2})
\end{equation}
where
\begin{equation}\label{eq:Pm-d3}
  C_{m}(k,\ell)
  = \sum_{j=1}^{n}\sum_{j'=1}^{n}
       [W^{m}]_{j,j'}\,
       e^{i(k\cdot y_{j'}- \ell\cdot y_{j}) },
\end{equation}
where $W^{m}$ is the $m$-th power of the matrix $W$,
with
\begin{equation}\label{eq:gamma-d3}
  W_{j,j'} =
  \begin{cases}
    \alpha_{j}, & j=j', \\[1ex]
    -\dfrac{e^{i\kappa |y_{j}-y_{j'}|}}{4\pi |y_{j}-y_{j'}|},
    & j\neq j'.
  \end{cases}
\end{equation}

In particular,
\begin{equation}\label{eq:P0-d3}
  \frac{-i}{2\pi^{2}}C_{0}(k,\ell)=f_{3,1}(k,\ell),\quad \frac{-2}{\pi} C_{1}(k,\ell)=f_{3,1}(k,\ell)+f_{3,2}(k,\ell),
\end{equation}
where $f_{3,1}$, $f_{3,2}$ are defined in (\ref{f3}).

\end{theorem}

Theorem \ref{high-energy-asymptotics} is proved in Section \ref{proof}. 

\begin{remark}
In dimension $d=1$ and $d=3$, the asymptotic expansions (\ref{eq:asymp-d1}) and (\ref{eq:asymp-d3}) converge to $f$.
\label{rk3.2}
\end{remark}

\begin{remark}
Our proof of Theorem \ref{high-energy-asymptotics} is based on asymptotic formulas (\ref{q(d=1)}), (\ref{q(d=2)}), (\ref{q(d=3)}) for $q$ in formulas (\ref{scattering amplitude}), (\ref{A=qb}). Substituting these asymptotic formulas for $q$ into formula (\ref{eigenfunctions}) for $\psi^{+}$, one gets high-energy asymptotic expansions for the scattering function $\psi^{+}$ for the case of multipoint potentials in (\ref{multipotential}).
\label{rk3.3} 
\end{remark}

We do not present in all detail the aforementioned full asymptotic expansions for $\psi^{+}$ at high energies because these formulas are similar to formulas in Theorem \ref{high-energy-asymptotics}. In the next result, we present only the leading terms of these asymptotic expansions.

\begin{theorem}[]\label{psi-asymptotics}
Let $\psi^{+}$ be the scattering solutions for a multipoint potential $v$ of the type (\ref{multipotential}). Then the leading asymptotics of $\psi^{+}$, as $\kappa := \sqrt{E} \to +\infty$ , are given by formula (\ref{eigenfunctions}) and the following formulas for $q$ : 

\medskip
If $d=1$, then
\begin{equation}
\label{q_main1}
     q_{j}(k)=-
       \,\alpha_{j}^{-1}\, e^{ik\cdot y_{j} } + \mathcal{O}(\kappa^{-1}) \quad \text{as } \kappa \to +\infty;
\end{equation}

\medskip
If $d=2$, then
\begin{equation}
\label{q_main2}
     q_{j}(k)= -2\pi(\ln \kappa)^{-1}\,e^{ik\cdot y_{j} }
      +4\pi^{2}(\ln \kappa)^{-2} 
        (\alpha_{j}-\frac{i}{4}) \,e^{ik\cdot y_{j} }
     + \mathcal{O}\!\bigl((\ln \kappa)^{-3}\bigr) \quad \text{as } \kappa \to +\infty;
\end{equation}

\medskip
If $d=3$, then
\begin{equation}
\label{q_main3}
     q_{j}(k)=  -i\,4\pi\,\kappa^{-1}
       e^{ik\cdot y_{j} }  - 16 \pi^{2}\,\kappa^{-2}  \left(\alpha_{j}\, e^{i k \cdot y_{j}}-\sum_{j' \neq j}\dfrac{1}{4\pi |y_{j}-y_{j'}|}\,
  e^{i\kappa |y_{j}-y_{j'}|} e^{ik\cdot y_{j'}}\right)
      + \mathcal{O}(\kappa^{-3}) \quad \text{as } \kappa \to +\infty. 
\end{equation}
\end{theorem}

Theorem \ref{psi-asymptotics} follows from formulas (\ref{eigenfunctions}), (\ref{q(d=1)}), (\ref{q(d=2)}), (\ref{q(d=3)}).

For multipoint potentials $v$ of Bethe-Peierls-Thomas-Fermi type, high-energy asymptotics of $\psi^{+}$ given in Theorem \ref{psi-asymptotics} can be already considered in some sense as analogs of formula (\ref{Yafayev}). 

In addition, in order to see the divergent beam transform $\mathcal{D}$ in asymptotics of $\psi^{+}$ in Theorem \ref{psi-asymptotics}, one can use formulas (\ref{g+beam_1}), (\ref{g+beam_2}) given below.

Let $g^{+}$ be defined by the formula
\begin{equation}
\label{g+}
   G^{+}(x,\kappa) = e^{ik\cdot x} g^{+}(x,k), \quad x, k \in \mathbb{R}^{d},\quad \kappa = |k|,
\end{equation}
where $G^{+}$ is defined by (\ref{Green}).

Then the following formulas hold:
\begin{align}
     \int_{\mathbb{R}^{d}} \, g^{+}(x-y,k)\, \phi(x) \, dx =  & (2 i \kappa) ^{-1}  \mathcal{D}\phi(y,\theta) + \mathcal{O}(\kappa^{-2}) \quad \text{as } \kappa \to  +\infty, \quad y \in \mathbb{R}^{d},  \label{g+beam_1}\\
     \mathcal{D}\phi(y,\theta) = & \int_{\mathbb{R}^{d}} \mathcal{D} \delta_{y} (x,-\theta)\, \phi(x)\, dx ,  \label{g+beam_2}
\end{align}
where $\phi$ is a compactly supported smooth test function on $\mathbb{R}^{d}$, $\delta_{y} =\delta (x-y)$, and $\delta$ is the standard Dirac delta function.

Formula (\ref{g+beam_1}) for $d=1$ easily follows from formula (\ref{Green}) for $d=1$. Formula (\ref{g+beam_1}) for $d \geq 2$ can be obtained proceeding from considerations given in \cite{buslaev1967trace}, \cite{yafaev2003high}, or, for example, proceeding from the proof of Proposition 3 in \cite{plamen1992generic}.

Proceeding from Theorem \ref{psi-asymptotics} and formulas (\ref{g+beam_1}), (\ref{g+beam_2}), we obtain the following result.

\begin{theorem}
\label{psi-beam transform}
    Let $\psi^{+}$ be the scattering solutions for a multipoint potential $v$ of the type (\ref{multipotential}). Then the following asymptotics formulas hold as $\kappa := \sqrt{E} \to +\infty$:

\medskip
If $d=1$, then
  \begin{equation}
\int_{\mathbb{R}} e^{-i k\cdot x} \, \psi^{sc}(x,k) \, \phi (x) \, dx  = (2i\kappa)^{-1} \int_{\mathbb{R}} \mathcal{D} v_{1,1} (x,-\theta)\, \phi(x)\, dx + \mathcal{O}(\kappa^{-2}) \quad \text{as } \kappa \to +\infty;
\label{yafayev_d=1}
  \end{equation}

If $d=2$, then
  \begin{align}
\int_{\mathbb{R}^{2}} e^{-i k\cdot x} \, \psi^{sc}(x,k) \, \phi (x) \, dx =&  (2i\kappa)^{-1}(\ln \kappa)^{-1}\int_{\mathbb{R}^{2}} \mathcal{D} v_{2,1} (x,-\theta)\, \phi(x)\, dx  \nonumber \\
+ & (2i\kappa)^{-1}(\ln \kappa)^{-2}\int_{\mathbb{R}^{2}} \mathcal{D} v_{2,2} (x,-\theta)\, \phi(x)\, dx  +
\mathcal{O}(\kappa^{-1}(\ln \kappa)^{-3}) \quad \text{as } \kappa \to +\infty ;
\label{yafayev_d=2}
  \end{align}

If $d=3$, then
  \begin{align}
\int_{\mathbb{R}^{3}} e^{-i k\cdot x} \, \psi^{sc}(x,k) \, \phi (x) \, dx =   (2i)^{-1} \kappa^{-2} \int_{\mathbb{R}^3} \mathcal{D} v_{3,1} (x,-\theta)\, \phi(x)\, dx + \mathcal{O}(\kappa^{-3}) \quad \text{as } \kappa \to +\infty.
\label{yafayev_d=3}
  \end{align}

Here, $\phi$ is a compactly supported smooth test function, and $v_{1,1}$, $v_{2,1}$, $v_{2,2}$, $v_{3,1}$ are defined in (\ref{delta11}), (\ref{delta21}), (\ref{delta22}), (\ref{delta31}).
\end{theorem}

One can see that the divergent beam transform arises in (\ref{yafayev_d=1})-(\ref{yafayev_d=3}) in a similar way with the "regular" formula (\ref{Yafayev}). 
Nevertheless, it should be noted that formula (\ref{yafayev_d=3}) shows only the leading asymptotic term which is independent on parameters $\alpha_{j}$'s in (\ref{multipotential}). 

Thus, for the case of multipoint potentials in (\ref{multipotential}), Remark \ref{rk3.3} and Theorems \ref{psi-asymptotics}, \ref{psi-beam transform} give analogs of formula (\ref{Yafayev}). However, for inverse scattering in this case, high-energy asymptotics of the scattering amplitude $f$ as in Theorem \ref{main asymptotics} seem to be much more convenient than those of $\psi^{sc}$. 

In addition, reductions of inverse scattering from boundary values of $\psi^{sc}$ to inverse scattering from $f$ are well known, see, for example, \cite{berezanskii1958uniqueness}; for such reductions as in  \cite{berezanskii1958uniqueness}, regularity or singularity of $v$ is not important.

\section{Proof of Theorem \ref{high-energy-asymptotics}} \label{proof}

\subsection{}Proof for $d=1$

For $d=1$, we present $A(\kappa)$ defined in (\ref{coefficientA}) as
\begin{equation}
    A(\kappa) = Z + G,
\end{equation}
where $Z$ and $G$ are defined by
\begin{equation}
    Z_{j,j'} := \alpha_{j} \delta_{j,j'},\quad
        G_{j,j'} := (2 i \kappa)^{-1}
      e^{i \kappa |y_{j}-y_{j'}|},
\end{equation}
where $\delta_{j,j'}$ is the Kronecker symbol. 
Thus,
\begin{align}
  A^{-1}(\kappa)
  &= \bigl(Z+G\bigr)^{-1}= \bigl(I + Z^{-1} G\bigr)^{-1}\,Z^{-1} \nonumber \\
  &= \sum_{m=0}^{M} \kappa^{-m}W^{m}\,Z^{-1}
     + \mathcal{O}(\kappa^{-M-1}) \quad \text{as } \kappa \to +\infty,\label{d=1:A-1}
\end{align}
where
\begin{equation}
   W=-\kappa Z^{-1}G,\quad W_{j,j'}=\left(\frac{i}{2}\right)\alpha_{j}^{-1}e^{i \kappa |y_{j}-y_{j'}|}. \label{d=1:Z-1G}
\end{equation}
Using formulas (\ref{A=qb}), (\ref{d=1:A-1}), (\ref{d=1:Z-1G}), we have

\begin{align}
  q_{j}(k)
  &= \sum_{j'=1}^{n}
       \bigl[A^{-1}(\kappa)\bigr]_{j,j'}\,
       (-e^{ik\cdot y_{j'} }) \nonumber \\
 &= -\sum_{m=0}^{M}\kappa^{-m}\, \sum_{j'=1}^{n}
       [W^{m}]_{j,j'}\,\alpha_{j'}^{-1}\,
       e^{ik\cdot y_{j'} } + \mathcal{O}(\kappa^{-M-1}) \quad \text{as } \kappa \to +\infty. \label{q(d=1)}
\end{align}
Substituting (\ref{q(d=1)}) into formula (\ref{scattering amplitude}) for $f(k,\ell)$, we get
\begin{align}
  f(k,\ell)
  &= \frac{1}{2\pi}
     \sum_{j=1}^{n} q_{j}(k)\,e^{- i \ell\cdot y_{j}} \nonumber \\
  &= \frac{-1}{2\pi} 
      \sum_{m=0}^{M}\kappa^{-m}\,\sum_{j=1}^{n}\sum_{j'=1}^{n}
       [W^{m}]_{j,j'}\,\alpha_{j'}^{-1}\,
       e^{i(k\cdot y_{j'}- \ell\cdot y_{j}) } + \mathcal{O}(\kappa^{-M-1})\quad \text{as } \kappa \to +\infty.
\end{align}
This completes the proof for $d=1$.

\subsection{}Proof for $d=2$

For $d=2$, we present $A(\kappa)$ defined in (\ref{coefficientA}) as
\begin{equation}
    A(\kappa) = \frac{\ln \kappa }{2\pi}\, I+Z + G,
\end{equation}
where $I$ is the identity matrix, and $Z$ and $G$ are defined by
\begin{equation}
    Z_{j,j'} =( \alpha_{j}  -\dfrac{i}{4} )\delta_{j,j'}
    ,\quad  G_{j,j} := 0,\quad
        G_{j,j'} := -\dfrac{i}{4}\,
       H_{0}^{(1)}\!\bigl(|y_{j}-y_{j'}|\kappa\bigr), \text{ if} \, j \neq j',
\end{equation}
where $\delta_{j,j'}$ is the Kronecker symbol. Thus,
\begin{align}
  A^{-1}(\kappa)
  &= \bigl(\frac{\ln \kappa}{2 \pi}\,I + Z+G\bigr)^{-1} = 2\pi(\ln \kappa)^{-1}
     \bigl(I + 2\pi (\ln \kappa)^{-1} Z
              + 2\pi (\ln \kappa)^{-1} G \bigr)^{-1} \nonumber \\
  &=2\pi(\ln \kappa)^{-1} \sum_{m=0}^{M} (-2\pi)^{m}(\ln \kappa)^{-m}
       Z^{m}
     + \mathcal{O}\!\bigl((\ln \kappa)^{-M-2}\bigr) \quad \text{as } \kappa \to +\infty.\label{d=2:A-1}
\end{align}
In (\ref{d=2:A-1}), we also use that
$$
  H_{0}^{(1)}(r) = \sqrt{\frac{2}{\pi}}\,e^{i(r-\frac{\pi}{4})} r^{-\frac{1}{2}}+ \mathcal{O}(r^{-1}) \quad \text{ as } \quad r \to +\infty,
$$
see, e.g., Section 9.2 of \cite{abramowitz1964}.
Using formulas (\ref{A=qb}), (\ref{d=2:A-1}), we have
\begin{align}
  q_{j}(k)
  &= \sum_{j'=1}^{n}
       \bigl[A^{-1}(\kappa)\bigr]_{j,j'}\,
       (-e^{ik\cdot y_{j'} }) \nonumber \\
  &=-2\pi(\ln \kappa)^{-1} \sum_{m=0}^{M} (-2\pi)^{m}(\ln \kappa)^{-m}
        \left(\alpha_{j}-\frac{i}{4}\right)^{m} \,e^{ik\cdot y_{j} }
     + \mathcal{O}\!\bigl((\ln \kappa)^{-M-2}\bigr) \quad \text{as } \kappa \to +\infty. \label{q(d=2)}
\end{align}
Substituting (\ref{q(d=2)}) into formula (\ref{scattering amplitude}) for $f(k,\ell)$, we get
\begin{align}
  f(k,\ell)
    &= (2\pi)^{-2}
     \sum_{j=1}^{n} q_{j}(k)\,e^{- i \ell\cdot y_{j}} \nonumber \\
  &= -(2 \pi \ln \kappa)^{-1} \sum_{m=0}^{M}
       (-2\pi)^{m} (\ln \kappa)^{-m}
       \sum_{j=1}^{n}
         \left(\alpha_{j}-\tfrac{i}{4}\right)^{m}\,
         e^{i (k-\ell)\cdot y_{j}} + \mathcal{O}\!\bigl((\ln \kappa)^{-M-2}\bigr) \quad \text{as } \kappa \to +\infty.
\end{align}
This completes the proof for $d=2$.

\subsection{}Proof for $d=3$

For $d=3$, we present $A(\kappa)$ defined in (\ref{coefficientA}) as
\begin{equation}
    A(\kappa) = \frac{-i}{4\pi}\,\kappa\, I + W,
\end{equation}
where $I$ is the identity matrix and $W$ is defined by
\begin{equation}
  W_{j,j}(\kappa) = \alpha_{j}, \qquad
  W_{j,j'}(\kappa) = -\dfrac{1}{4\pi |y_{j}-y_{j'}|}\,
  e^{i\kappa |y_{j}-y_{j'}|}, \quad j\neq j'.
\end{equation}
Thus,
\begin{align}
  A^{-1}(\kappa)
  &= \bigl(\frac{-i}{4\pi}\,\kappa\, I + W\bigr)^{-1} = i\,4\pi\,\kappa^{-1}
     \Bigl(I + i\,4\pi\,\kappa^{-1}W\Bigr)^{-1} \nonumber \\
  &= i\,4\pi\,\kappa^{-1}\sum_{m=0}^{M}
       (-i\,4\pi)^{m}\,
       \kappa^{-m} W^{m}
     + \mathcal{O}(\kappa^{-M-2}) \quad \text{as } \kappa \to +\infty\label{d=3:A-1}.
\end{align}
Using formulas (\ref{A=qb}), (\ref{d=3:A-1}), we have
\begin{align}
  q_{j}(k)
  &= \sum_{j'=1}^{n}
       \bigl[A^{-1}(\kappa)\bigr]_{j,j'}\,
       (-e^{ik\cdot y_{j'} }) \nonumber \\
 &= -i\,4\pi\,\kappa^{-1}\sum_{m=0}^{M}(-i\,4\pi)^{m}\,\kappa^{-m}\, \sum_{j'=1}^{n}
       [W^{m}]_{j,j'}\,
       e^{ik\cdot y_{j'} } + \mathcal{O}(\kappa^{-M-2}) \quad \text{as } \kappa \to +\infty. \label{q(d=3)}
\end{align}

Substituting (\ref{q(d=3)}) into formula (\ref{scattering amplitude}) for $f(k,\ell)$, we get
\begin{align}
  f(k,\ell)
  &= (2\pi)^{-3}
     \sum_{j=1}^{n} q_{j}(k)\,e^{- i \ell\cdot y_{j}} \nonumber \\
  &= -\frac{i}{2\pi^2}\,\kappa^{-1}
      \sum_{m=0}^{M}(-i\,4\pi)^{m}\,\kappa^{-m}\,\sum_{j=1}^{n}\sum_{j'=1}^{n}
       [W^{m}]_{j,j'}\,
       e^{i(k\cdot y_{j'}- \ell\cdot y_{j}) } + \mathcal{O}(\kappa^{-M-2})\quad \text{as } \kappa \to +\infty.
\end{align}
This completes the proof for $d=3$.

\section*{Acknowledgements}
This work was started during the internship of the first author in the Centre de Mathématiques Appliquées of École polytechnique in February-May 2025.

\bibliographystyle{plain}     
\bibliography{scatter}   

@article{kuo2024inverse,
  title={Inverse scattering for the multipoint potentials of {B}ethe- {P}eierls- {T}homas- {F}ermi type},
  author={P.C. Kuo and R.G. Novikov},
  journal={Inverse Problems},
  volume = {41},
  number = {6},
  pages = {065021},
  year={2025},
  publisher = {IOP Publishing},
}

@article{grinevich2021transmission,
  title={Transmission eigenvalues for multipoint scatterers},
  author={P.G. Grinevich and R.G. Novikov},
  journal={Eurasian Journal of Mathematical and Computer Applications},
  volume={9},
  number={4},
  pages={17--25},
  year={2021}
}

@book{abramowitz1964,
  title     = {Handbook of Mathematical Functions with Formulas, Graphs, and Mathematical Tables},
  editor    = {M. Abramowitz and I.A. Stegun},
  year      = {1964},
  publisher = {National Bureau of Standards},
  address   = {Washington, DC},
  series    = {Applied Mathematics Series},
  volume    = {55}
}

@book{albeverio2012solvable,
  title={Solvable models in quantum mechanics, Texts and Monographs in Physics},
  author={S. Albeverio and F. Gesztesy and R. Hegh-Krohn and H. Holden},
  year={1998},
  publisher={Springer-Verlag, New York}
}

@article{bethe1935quantum,
  title={Quantum theory of the diplon},
  author={H. Bethe and R. Peierls},
  journal={Proceedings of the Royal Society of London. Series A-Mathematical and Physical Sciences},
  volume={148},
  number={863},
  pages={146--156},
  year={1935},
  publisher={The Royal Society London}
}

@inproceedings{novikov2019multidimensional,
  title={Multidimensional inverse scattering for the {S}chr{\"o}dinger equation},
  author={R.G. Novikov},
  booktitle={ISAAC Congress (International Society for Analysis, its Applications and Computation)},
  pages={75--98},
  year={2019},
  organization={Springer}
}

@article{born1926quantenmechanik,
  title={Quantenmechanik der sto{\ss}vorg{\"a}nge},
  author={M. Born},
  journal={Zeitschrift f{\"u}r physik},
  volume={38},
  number={11},
  pages={803--827},
  year={1926},
  publisher={Springer}
}

@article{thomas1935interaction,
  title={The interaction between a neutron and a proton and the structure of {$H^3$}},
  author={L.H. Thomas},
  journal={Physical review},
  volume={47},
  number={12},
  pages={903--909},
  year={1935},
  publisher={APS}
}

@article{fermi1936sul,
  title={Sul moto dei neutroni nelle sostanze idrogenate},
  author={E. Fermi},
  journal={Ricerca scientifica},
  volume={7},
  number={2},
  pages={13--52},
  year={1936}
}

@article{badalyan2009scattering,
  title={Scattering by acoustic boundary scatterers with small wave sizes and their reconstruction},
  author={N.P. Badalyan and V.A. Burov and S.A. Morozov and O.D. Rumyantseva},
  journal={Acoustical Physics},
  volume={55},
  pages={1--7},
  year={2009},
  publisher={Springer}
}

@article{dmitriev2021features,
  title={Features of solving the direct and inverse scattering problems for two sets of monopole scatterers},
  author={K.V. Dmitriev and O.D. Rumyantseva},
  journal={Journal of Inverse and Ill-posed Problems},
  volume={29},
  number={5},
  pages={775--789},
  year={2021},
  publisher={De Gruyter}
}

@inproceedings{grinevich2024transparent,
  title={Transparent scatterers and transmission eigenvalues},
  author={P.G. Grinevich and R.G. Novikov},
  booktitle={Inverse Problems: Modelling and Simulation. Trends in Mathematics
  },
  volume={11},
  pages={265--274},
  year={2025},
  organization={ Birkhäuser, Cham}
}

@article{grinevich2012faddeev,
  title={Faddeev eigenfunctions for multipoint potentials},
  author={P.G. Grinevich and R.G. Novikov},
  journal={Eurasian Journal of Mathematical and Computer Applications},
  volume={1},
  number={2},
  pages={76--91},
  year={2013}
}

@article{agaltsov2019examples,
  title={Examples of solution of the inverse scattering problem and the equations of the {N}ovikov--{V}eselov hierarchy from the scattering data of point potentials},
  author={A.D. Agaltsov and R.G. Novikov},
  journal={Russian Mathematical Surveys},
  volume={74},
  number={3},
  pages={373--386},
  year={2019},
  publisher={IOP Publishing}
}

@article{novikov2018inverse,
  title={Inverse scattering for the {B}ethe-{P}eierls model},
  author={R.G. Novikov},
  journal={Eurasian Journal of Mathematical and Computer Applications},
  volume={6},
  number={1},
  pages={52--55},
  year={2018}
}

@article{Berezin:1960df,
    author = "F.A. Berezin and L.D. Faddeev",
    title = "{A remark on {S}chr{\"o}dinger's equation with a singular potential}",
    journal = "Soviet Math. Dokl. ",
    volume = "2",
    pages = "372--375",
    year = "1961"
}

@article{mantile2023inverse,
  title={Inverse wave scattering in the time domain for point scatterers},
  author={A. Mantile and A. Posilicano},
  journal={Journal of Mathematical Analysis and Applications},
  volume={518},
  number={2},
  pages={126758},
  year={2023},
  publisher={Elsevier}
}

@article{gesztesy1999inverse,
  title={An inverse problem for point inhomogeneities},
  author={F. Gesztesy and A.G. Ramm},
  journal={Methods of Functional Analysis and Topology},
  volume={6},
  number={2},
  pages={1--12},
  year={1999}
}

@article{grinevich2022spectral,
  title={Spectral inequality for {S}chr{\"o}dinger's equation with multipoint potential},
  author={P.G. Grinevich and R.G. Novikov},
  journal={Russian Mathematical Surveys},
  volume={77},
  number={6},
  pages={1021--1028},
  year={2022},
  publisher={Russian Academy of Sciences, Steklov Mathematical Institute of Russian~…}
}

@book{demkov2013zero,
  title={Zero-range potentials and their applications in atomic physics},
  author={Y.N. Demkov and V.N. Ostrovskii},
  year={2013},
  publisher={Springer New York, NY}
}

@article{malamud2024kernels,
  title={On kernels of invariant {S}chr{\"o}dinger operators with point interactions. {G}rinevich--{N}ovikov problem},
  author={M.M. Malamud and V.V. Marchenko},
  journal={Doklady Mathematics},
  volume={109},
  number={2},
  pages={125--129},
  year={2024}
}

@article{loran2022renormalization,
  title={Renormalization of multi-delta-function point scatterers in two and three dimensions, the coincidence-limit problem, and its resolution},
  author={F. Loran and A. Mostafazadeh},
  journal={Annals of Physics},
  volume={443},
  pages={168966},
  year={2022},
}

@article{chashchin2018example,
  title={Example of point potential with inner structure},
  author={D.S. Chashchin},
  journal={Eurasian Journal of Mathematical and Computer Applications},
  volume={6},
  number={1},
  pages={4--10},
  year={2018},
}

@article{kurasov2009triplet,
  title={Triplet extensions {I}: {S}emibounded operators in the scale of {H}ilbert spaces},
  author={P. Kurasov},
  journal={Journal d'Analyse Mathematique},
  volume={107},
  number={1},
  pages={251--286},
  year={2009},
}

@article{yafaev2003high,
  title={High-energy and smoothness asymptotic expansion of the scattering amplitude},
  author={D. Yafaev},
  journal={Journal of Functional Analysis},
  volume={202},
  number={2},
  pages={526--570},
  year={2003},
  publisher={Elsevier}
}

@article{faddeev1956uniqueness,
  title={The uniqueness of solutions for the scattering inverse problem},
  author={L.D. Faddeev},
  journal={Vestnik Leningrad Univ},
  volume={7},
  pages={126--130},
  year={1956}
}

@article{novikov1996inverse,
  title={Inverse scattering up to smooth functions for the {S}chr{\"o}dinger equation in dimension 1},
  author={R.G. Novikov},
  journal={Bulletin des sciences math{\'e}matiques (Paris. 1885)},
  volume={120},
  number={5},
  pages={473--491},
  year={1996}
}

@book{natterer2001mathematics,
  title={The mathematics of computerized tomography},
  author={F. Natterer},
  year={2001},
  publisher={SIAM}
}

@article{ie1957behavior,
  title={On the behavior of eigenfunctions of the {S}chr{\"o}dinger equation},
  author={\`E.~\`E.~Shnol'},
  journal={PhD thesis},
  volume={Moscow State University},
  pages={},
  year={1955}
}

@article{buslaev1967trace,
  title={Trace formulas and certain asymptotic estimates of the resolvent kernel for the {S}chr{\"o}dinger operator in three-dimensional space},
  author={V.S. Buslaev},
  journal={Topics in Math. Phys},
  volume={1},
  year={1967},
  publisher={Plenum Press, Oxford}
}

@article{berezanskii1958uniqueness,
  author  = {Y.M. Berezanskii},
  title   = {The uniqueness theorem in the inverse problem of spectral analysis for the {S}chr{\"o}dinger equation},
  journal = {Tr. Mosk. Mat. Obshch.},
  volume  = {7},
  pages   = {3--62},
  year    = {1958 (in Russian)},
  note    = {English translation: Transl., Ser. 2, Am. Math. Soc. 35: 167--235, 1964.}
}

@article{plamen1992generic,
  title={Generic uniqueness for two inverse problems in potential scattering},
  author={P. Stefanov},
  journal={Communications in partial differential equations},
  volume={17},
  number={1-2},
  pages={55--68},
  year={1992},
  publisher={Taylor \& Francis}
}
\newpage
\vspace{2cm}
\noindent
Pei-Cheng Kuo\\
Université de Versailles Saint-Quentin-en-Yvelines, Université Paris-Saclay,\\
 45 Av. des États Unis, 78000 Versailles, France;\\
E-mail: pei-cheng.kuo@ens.uvsq.fr 
\\
\\

\noindent
Roman G. Novikov\\
CMAP, CNRS, École polytechnique,\\
Institut polytechnique de Paris, 91120 Palaiseau, France;\\
E-mail: novikov@cmap.polytechnique.fr

\end{document}